\documentclass[%
 aip,
 amsmath,amssymb,
 reprint,%
]{revtex4-1}

\usepackage{graphicx}
\usepackage{dcolumn}
\usepackage{bm}

\usepackage[utf8]{inputenc}
\usepackage[T1]{fontenc}
\usepackage{mathptmx}

\begin{document}

\preprint{AIP/123-QED}

\title[New algebraically solvable systems\ldots]{New algebraically solvable systems of two 
autonomous first-order ordinary differential equations with purely quadratic
right-hand sides}

\author{F. Calogero}
 \altaffiliation[Also at ]{Istituto Nazionale di Fisica Nucleare, Sezione di Roma, Italy.}

 \email{francesco.calogero@roma1.infn.it, francesco.calogero@uniroma1.it.}
\affiliation{ 
Dipartimento di Fisica, Universit\`{a} di Roma ``La
Sapienza'', 00185 Roma, Italy.
}

\author{R. Conte}
 \email{robert.conte@cea.fr}

\affiliation{%
Universit\'e Paris-Saclay, ENS Paris-Saclay, CNRS, Centre Borelli, F-94235, Cachan, France;
Department of mathematics, The University of Hong Kong, Pokfulam
road, Hong Kong.
}%

\author{F. Leyvraz}
 \altaffiliation[Also at ]{Centro Internacional de Ciencias, Cuernavaca, Morelos M\'exico}

\email{f\_leyvraz2001@hotmail.com}

\affiliation{%
Instituto de Ciencias F\'\i sicas---Universidad Nacional Aut\'onoma de
M\'exico\\ 
Cuernavaca, 62210 Morelos, M\'exico.
}

\date{\today}

\begin{abstract}
We identify many  new solvable  subcases of the general dynamical
system characterized by  two   autonomous   first-order ordinary
differential equations  with  purely quadratic  right-hand sides; the 
 solvable  character of these dynamical systems amounting to the
possibility to obtain the solution of their initial value problem via 
 algebraic  operations. Equivalently---by considering the analytic
continuation of these systems to  complex  time---their  
algebraically solvable  character corresponds to the fact that their 
general  solution is either  singlevalued  or features only a  
finite number  of  algebraic branch points  as functions of  
complex  time (the independent variable). Thus our results provide a major
enlargement of the class of  solvable  systems beyond those with 
singlevalued  general solution identified by Garnier about 60 years
ago. An interesting property of several of these new dynamical systems is
the elementary character of their general solution, identifiable as the roots of
a polynomial with explicitly obtainable time-dependent coefficients. We also
mention that, via a well-known  time-dependent  change of (dependent
and independent) variables featuring the  imaginary  parameter $%
\mathbf{i} \omega $ (with $\omega $ an  arbitrary strictly positive real 
number),  autonomous  variants can be explicitly exhibited of each of
the  algebraically solvable  models we identify: variants which 
 all  feature the remarkable property to be  isochronous, 
 i.e. their  generic  solution is periodic  with a
period that is a  fixed integer multiple  of the basic period 
$T=2\pi/\omega$.
\end{abstract}

\maketitle

\newcommand{\sign}{\mbox{sgn}\,}

\section{Introduction}
\label{sec:intro}

Two general approaches can be distinguished in the investigation of
dynamical systems, which in this paper are identified as autonomous
systems of ordinary differential equations (ODEs) with, generally %
nonlinearly coupled, right-hand sides; of course beyond general theorems
about the existence and uniqueness of solutions to  such problems. One point of
view---which might perhaps be symbolically characterized with the name of
Painlev\'{e}---focusses on the identification of such systems which can, in
some sense, be characterized as solvable: of course by restricting on
a case-by-case manner the techniques to be used for their solution, or the
mathematical properties to be satisfied by their general solution. A complementary
point of view---which might perhaps be symbolically identified with the name
of Poincar\'{e}---focusses instead on the identification of specific,
interesting, features of the solution of given systems, such as
the existence of equilibria and the behavior in their vicinity, asymptotic
properties, the sensitivity of the dependence on initial data and related
issues having to do with the notion of deterministic chaos.

Clearly these two  points of view are complementary, while the approaches they
entail are quite different: in the first (``Painlev\'e'') case the goal is
to identify special dynamical systems---possibly within certain
general classes---which feature some special property, in
particular that to be in some sense solvable (see below); in the
second (``Poincar\'{e}'') case the goal is to investigate special
properties of some given dynamical system.

In this paper we adopt the first (``Painlev\'e'') point of view, and focus
on what might be considered the prototypical class of nonlinear dynamical systems: we
study the following system of ODEs: 
\begin{subequations}
\label{general}
\begin{eqnarray}
\dot{x}_{1}&=&c_{11}x_{1}^{2}+c_{12}x_{2}^{2}+c_{13}x_{1}x_{2},
\label{generala}
\\
\dot{x}_{2}&=&c_{21}x_{1}^{2}+c_{22}x_{2}^{2}+c_{23}x_{1}x_{2}.
\label{generalb}
\end{eqnarray}
\end{subequations}
\textbf{Notation 1.1.} Hereafter $t$ (``time'') is the independent variable,
superimposed dots denote $t$-differentiations, we generally use the last
letters of the Latin alphabet (such as $x,$ $y,$ $z;$ possibly equipped with
indices) to denote time-dependent variables (generally without
indicating explicitly their time-dependence: hence for instance $%
x_{1}\equiv x_{1}\left( t\right) $, $\dot{x}_{2}\equiv dx_{2}\left( t\right)
/dt$), and the first letters (such as $a,$ $b,$ $c$, $A,$ $B,$ $%
C;$ possibly equipped with indices) of the Latin alphabet to denote
time-independent quantities, such as parameters (as, say, $c_{12}$:
other time-independent quantities are of course the initial values,
such as, say, $x_{1}\left(0\right)$). Generally all quantities are 
complex numbers (and $\mathbf{i}$ denotes the imaginary unit: $\mathbf{i}%
^{2}=-1$); except for the time $t$ which is generally the real
independent variable (although consideration of the analytic
continuation of the dependent variables to complex values
of the independent variable $t$ shall also turn out to be quite
useful in the following treatment: see below). $\blacksquare $

The main results of this paper are the following: we define a concept 
of algebraic solvability, see below, and identify a large number
of systems of the form (\ref{general}) having this property. We 
further show its importance, both because it allows us
to display the systems' solutions in a rather explicit  manner, and 
also because it allows to associate to each algebraically 
solvable system of the form (\ref{general}) a related system 
with added linear terms, see (\ref{isogen}) below, with the 
property that all its solutions are periodic with the same period,
except possibly for an exceptional set of orbits passing through a 
singularity.

We define an algebraically solvable system as having the
property that their solutions feature at most finitely many
\textit{algebraic} branch point singularities, so that they are defined, for
complex times, on a Riemann surface with a \textit{finite} number of
sheets. This property, which is central to the paper, extends the 
class studied by Garnier \cite{garnier1, garnier2}, who identified all systems of two homogeneous 
ODE's having the property that their solutions, as
functions of complex times, are singlevalued, in other words, that
they have no branch point singularities.
In the following, we identify a countable infinity of
systems of quadratic ODE's with 
the more general property of algebraic solvability, and show that their solutions
can be displayed in an explicit manner. 

Let us illustrate our meaning by showing an instance of such an 
algebraically solvable system: let us consider the special case of 
(\ref{general}) given by 
\begin{subequations}
\label{eqsspecial}
\begin{eqnarray}
\dot{x}_{1}&=&x_{1}x_{2}~,  
\label{eqsspeciala}
\\
\dot{x}_{2}&=&\left( 3/4\right) \left( x_{1}^{2}+x_{2}^{2}\right) ~.
\label{eqsspecialb}
\end{eqnarray}%
It follows from our results (see below {\bf Section \ref{sec:solution}}), that it has the
following general solution: 
\end{subequations}
\begin{subequations}
\label{solex}
\begin{eqnarray}
x_{1}(t)&=&\frac{9C}{\left[ 3-w(t)^{2}\right] ^{2}},  \label{solexa} \\
x_{2}(t)&=&\frac{9Cw(t)}{\left[ 3-w(t)^{2}\right] ^{2}},  \label{solexb} \\
\frac{w(t)^{3}}{9}-w(t)+\frac{3C}{4}(t-t_0)&=&0,  \label{solexc}
\end{eqnarray}%
with $C$ and $t_0$ given in terms of the initial data by the following
formulas: 
\end{subequations}
\begin{subequations}
\begin{eqnarray}
C &=&\frac{\left[ 3x_{1}(0)^{2}-x_{2}(0)^{2}\right] ^{2}}{9x_{1}(0)^{3}},
\label{C} \\
t_0 &=&\frac{4x_{2}(0)\left[ x_{2}(0)^{2}-9x_{1}(0)^{2}\right] }{3%
\left[ x_{2}(0)^{2}-3x_{1}(0)^{2}\right] ^{2}};  \label{tbar}
\end{eqnarray}%
and also note (see (\ref{solexa}) and (\ref{solexb})) that, for all
values of $t$, 
\end{subequations}
\begin{equation}
w\left( t\right) =x_{2}\left( t\right) /x_{1}\left( t\right) ~.  \label{w}
\end{equation}

These formulas provide quite  explicitly the solution of
the initial value problem for the system (\ref{eqsspecial}), except for the
relatively minor---but quite significant---problem of identifying which one
of the  three  solutions $w\left( t\right) $ of the cubic equation (\ref{solexc}) 
should be inserted in the right-hand side of the  two  formulas 
(\ref{solexa}) and (\ref{solexb}). The way to address this issue is 
described in detail in \cite{FC1966}.

Let us emphasize that this example shows that the solution of the
initial value problem for the system of nonlinearly-coupled ODEs 
(\ref{eqsspecial}) has been reduced to a purely algebraic problem: in
this particular case, to the quite simple one of finding the roots of a 
cubic equation, a problem that can even be solved explicitly
via the well-known Cardano formulas. This kind of reduction of the
solution of the initial value problem for the dynamical system under
consideration to algebraic operations is the characteristic feature
of the subclass of the dynamical system (\ref{general}) which are identified in
this paper; which should therefore be characterized as  {\em algebraically
solvable}, although for brevity we often use the term  {\em solvable.}
We further note that the simplicity of the solution of
(\ref{eqsspecial}) described above, generally arises whenever, 
as in this example, the Riemann surface on which the ODE's solution is defined,
has finitely many sheets. This condition implies, of course, both that
all branch points be algebraic and that there be only finitely many such branch points. 

 Let us briefly compare our approach to others. Apart from qualitative 
and geometric approaches,
which are not readily applicable to systems of two complex ODE's, an important technique
for the study of the kind of polynomial ODE's we address, has been the search for polynomial,
and more generally speaking, algebraic, invariants, according to the approach initially
pioneered by Darboux \cite{darboux}. Searching for such invariants has been done for 
systems similar to ours, such as the Lotka --Volterra system. This is the special case of (\ref{general})
in which $c_{12}=c_{21}=0$, but for which linear growth or decay 
terms are additionally taken into account, see for example \cite{invariant}.
The existence of such invariants can be related to the value
of a set of explicitly computable quantities known as the Kowalevski exponents, see 
\cite{kowal, yoshida1, yoshida2}. The cases for which the Kowalewski invariants
are all integers have been determined in \cite{SW}. For these it can be shown that 
the corresponding solutions are singlevalued in the plane, and that they therefore correspond
to the class of solutions investigated by Garnier.
The relation of our approach to the techniques involving the search for invariants
is not clear to us. We assume our approach may be more general.
Certainly it allows to address directly the problem of determining the time evolution explicitly, 
which does not follow immediately from the knowledge of an invariant quantity.

In order to proceed, it is convenient to introduce a 
canonical form of the system (\ref{general}), reading as follows:
\begin{subequations}
\label{nor}
\begin{eqnarray}
\dot{x}_{1}&=&x_{1}x_{2},  \label{norma}
\\
\dot{x}_{2}&=&A(x_{1}^{2}+x_{2}^{2})+Bx_{1}x_{2},  \label{normb}
\end{eqnarray}
\end{subequations}
which features in the right-hand side of its ODEs only the  two  parameters $A$
and $B$ (rather than the  six  parameters $c_{nj},$ see (\ref{general})). This
can be done (as explained in {\bf Section \ref{sec:canon}}) using the obvious possibility to
perform a linear transformation (featuring  four {\em a priori}
arbitrary parameters) of the  two   dependent variables; a change 
which modifies only quite marginally the nature of the
problem. It should be pointed out that there exist exceptional 
cases for which the reduction to (\ref{nor}) cannot be performed.
These can be reduced to a number of simpler forms 
involving only one parameter, see {\bf Section \ref{sec:canon}}.

\textbf{Remark 1.1}. Note that, since the systems (\ref{general}) and (\ref%
{nor}) generally involve complex variables, they may also be viewed
as a system of  four  real ODE's, involving the real and imaginary parts of $x_1$ 
and $x_2$. It is therefore not immediately accessible to
standard qualitative approaches valid for  two-dimensional real
systems. $\blacksquare $

This canonical  form allows us to discuss the classical
investigation of the system (\ref{general}), as made about 60 years ago
by Garnier \cite{garnier1,garnier2}. The focus of those studies is indeed
on the \textit{analyticity} properties of the solution of the system 
(\ref{general}) when considered as functions of the complex variable $t$,
and it led to the identification of all
the subsystems of (\ref{general}) such that their general solution
is a singlevalued function of the complex variable $t$: either they are 
entire functions of $t$---i.e., free of singularities in the entire
complex $t$-plane---or their only singularities are  
unramified. This restriction amounts to the exclusion of all dynamical
systems (\ref{general}) the solution of which features branch points
in the complex $t$-plane. And Garnier \cite{garnier1,garnier2} was
able to identify {\em all\/} the dynamical systems (\ref{general}) of this
restricted kind. For the canonical form (\ref{nor}), the
systems found by Garnier \cite{garnier1, garnier2} correspond to the
trivially solvable
case $A=0$ and $B$ arbitrary, as well as to  five  additional cases in which
both the values of $A$ and $B$ take well-defined values (see below). In this
sense, there are only a small number of systems of that type.

The main results of the present paper consist in the introduction and
identification of a much larger subclass of the dynamical systems (\ref%
{general}), of which the system (\ref{eqsspecial}) discussed just above is a
simple example. These systems---which can be identified as ``solvable'' (or
``algebraically solvable'': see above)---might be characterized (to make
contact with Garnier's approach)---as the dynamical systems (\ref{general})
the general solution of which---as functions of the  complex
variable $t$---defines a Riemann surface with a \emph{finite\/}
number of sheets. This implies, of course, that it can only feature a 
finite number of  algebraic branch points (i.e.,
singularities of type $\left( t-t_{S}\right) ^{p}$ with $p$ restricted to be
a rational number: as implied by the above discussion, the example (%
\ref{eqsspecial}) discussed above is such an example in which both
singularities with $p=1/3$ and $p=1/2$ arise).

In the following we show---by discussing in detail how the solution of these
dynamical systems can be obtained---that it is indeed justified to
categorize these systems as  solvable; and that this entails a very
substantial enlargement of the  solvable subclass of dynamical
systems (\ref{general}), beyond the class previously identified by Garnier 
\cite{garnier1, garnier2}.

These findings are important because obviously the class of nonlinear
dynamical systems (\ref{general}) is quite interesting, both due to its
mathematical neatness and even more so because of its relevance in many
applicative contexts.

There is an additional reason why the solvable subclass of systems
we identify is particularly interesting. This is a consequence of the following
observation (see for instance \cite{FC2008} and references therein): let the system
(\ref{general}) be reexpressed in terms of a new independent variable
$\tau$, as follows (note that this is the only place in the paper where we shall use
an independent variable different from $t$): 
\begin{subequations}
\label{generaltau}
\begin{eqnarray}
\frac{dx_1}{d\tau}&=&c_{11}x_{1}^{2}+c_{12}x_{2}^{2}+c_{13}x_{1}x_{2},
\label{generaltaua}
\\
\frac{dx_1}{d\tau}&=&c_{21}x_{1}^{2}+c_{22}x_{2}^{2}+c_{23}x_{1}x_{2}.
\label{generaltaub}
\end{eqnarray}
\end{subequations}
If one now performs the following simple change of dependent and 
independent variables, 
\begin{subequations}
\label{isotrans}
\begin{eqnarray}
x_{1}(t)&=&\exp \left( -\mathbf{i}\omega t\right) \tilde{x}_1\left( \tau
\right) ,\\
 \label{isotransa}
x_{2}(t)&=&\exp \left( -\mathbf{i}\omega t\right)
\tilde{x}_{2}\left( \tau \right) ,  \label{isotransb}
\\
\tau&\equiv&\tau \left( t\right) =\frac{\exp \left( \mathbf{i}\omega
t\right) -1}{\mathbf{i}\omega },  \label{isotransc}
\end{eqnarray}
\end{subequations}
the autonomous system (\ref{generaltau}) gets transformed into the
following, also autonomous, system: 
\begin{subequations}
\label{isogen}
\begin{eqnarray}
\dot{\tilde{x}}_{1}&=&\mathbf{i}\omega \tilde{x}_{1}+c_{11} \tilde{x}%
_{1}  ^{2}+c_{12}\tilde{x}_{2}  ^{2}+c_{13}\tilde{x}_{1}%
\tilde{x}_{2},
\\
\dot{\tilde{x}}_{2}&=&\mathbf{i}\omega \tilde{x}_{2}+c_{21} \tilde{x}%
_{1}  ^{2}+c_{22}\tilde{x}_{2}  ^{2}+c_{23}\tilde{x}_{1}%
\tilde{x}_{2}.
\end{eqnarray}
\label{QuadtIso}
\end{subequations}
Here and hereafter $\omega $ is a strictly positive real parameter; note
that its presence multiplied by the imaginary unit $\mathbf{i}$
implies that the dependent variables $\tilde{x}_{1}\equiv \tilde{x}%
_{1}\left( t\right) $ and $\tilde{x}_{2}\equiv \tilde{x}_{2}\left( t\right) $
evolving according to this system of ODEs take necessarily complex
values---so that this system might equivalently be considered a system of  four 
real dependent variables; while here we mainly restrict attention
to real values of the independent variable $t$ (``time'').

It is then obvious that the time evolution of this system, (\ref{isogen}),
corresponds in a quite straightforward manner to the evolution of the system
(\ref{generaltau}) when the  complex time $\tau$
of that system, see (\ref{isotransc}),  goes
round and round, counterclockwise, on the circle of radius $1/\omega$ 
centered at the point $\mathbf{i}/\omega $ 
in the complex $\tau $-plane. This implies---rather
obviously, in the context of the results described above (or see, if need
be, \cite{FC2008, GS2005})---that the time evolution of the
system (\ref{isogen})---if this system is obtained from one of the 
solvable systems identified in this paper, see below---features the highly
remarkable property to be  isochronous, namely its general
solution is  periodic with a period independent of the initial conditions
and given by an integer multiple
of the basic period $T=2\pi /\omega$.  Note that this need not
hold for strictly all orbits, since the circle on which $\tau$ moves may hit a branch point
singularity, after which the solution is not uniquely defined any more.
In fact, at the boundary between two regions in which the solution has {\em different\/}
periods $n_1T$ and $n_2T$, with $n_1$ and $n_2$ different integers, this must necessarily happen.

The existence of a common period for all orbits, given by $mT$ with
$m$ a fixed strictly positive integer, rests essentially 
on the property that the Riemann surface describing the solution only
has a finite number of sheets, so that the evolution of $\tau$ on the
circle always eventually returns to the sheet from which it started. 
For more general systems, it may happen, for instance, that every orbit is periodic with
a period $nT$, with $n\in\mathbb{N}$, but that $n$ can take arbitrarily high 
values, depending on the initial condition. In other cases, it
may happen that asymptotically periodic orbits, as for instance
described in \cite{AIS}, or
more general aperiodic orbits \cite{GS2007}  are generated.

Let us end this introductory survey of the results reported in this
paper---which has been mainly meant to illustrate what we mean by the
statement that a certain dynamical system is solvable---with a
final observation, introduced in order to eliminate a possible
misunderstanding. It is sometimes stated that a system of ODEs is  
{\em solvable by quadratures}; this is indeed the case for the system (%
\ref{general}), as is well-known, see for example \cite{nick} and below.
This actually means that the time $t$ can be expressed via integrals
involving the functions $x_{1}\left( t\right) $ and $x_{2}\left( t\right) $
in such a way that the determination of these variables as a function of
time is achieved by \textit{inverting} these relations:
see below. This task, however, can be relatively simple or exceedingly
complicated; so that in the first case this approach does indeed allow to
obtain the solution of the dynamical system (\ref{general})---i.e., to
determine for given initial values $x_{1}\left( 0\right) $ and $x_{2}\left(
0\right) $ the solutions $x_{1}\left( t\right) $ and $x_{2}\left( t\right) $
for all values of $t>0$ or even for complex values of $t$;
while in the second case it would be hardly justified to assert that the 
solution of the dynamical system (\ref{general}) can be  
obtained by quadratures. The distinction among these two
alternative situations shall become more clear from the treatment given
below. Our point of view is that only in the first case the solution of the
dynamical system (\ref{general}) has been achieved; indeed our approach will
be to analyze the solution by  quadratures and to identify those
special cases in which the  inversion reduces to an 
algebraic procedure.

As an additional result, we have shown that the case of (\ref{nor}) with $B=0$ can be
reduced to a particular case of the
one-dimensional Newton equation (``acceleration equals force''), thus
allowing some further results to be derived. In
particular it is shown there that the very simple {\em complex} Newtonian equation 
\begin{subequations}
\label{SimpleNewt}
\begin{equation}
\ddot{\zeta}=\zeta ^{k}  \label{SimNew}
\end{equation}%
is algebraically solvable for the following infinite sets
of values of the parameter $k$ 
\begin{equation}
k=-\left( 2n+1\right) /\left( 2n-1\right) \text{ and }k=-\left( n-1\right)/n,
\label{kknn}
\end{equation}
\end{subequations}
for $n\in\mathbb{N}$.
And in particular this implies that
the following {\em complex\/} variant of the Newtonian equations (\ref{SimpleNewt}), 
\begin{equation}
\ddot{\tilde{\zeta}}=\mathbf{i}\left( \frac{k+3}{k-1}%
\right) \omega \dot{\tilde{\zeta}}+\left[ \frac{2\left(
k+1\right) }{\left( k-1\right) ^{2}}\right] \omega ^{2}\tilde{\zeta}+\tilde{%
\zeta}^{k}
\label{NewtIso}
\end{equation}%
(with $\omega $ an arbitrary strictly positive parameter, $\mathbf{i}$
the imaginary unit, and $k$ given by one of the values
of (\ref{kknn})), is  isochronous for real values of $t$. 

In \textbf{Section \ref{sec:canon}} we set up the problem in some detail and describe the
results obtained by Garnier \cite{garnier1, garnier2}. In 
\textbf{Section \ref{sec:solution}} we describe our general approach. 
In \textbf{Section \ref{sec:newton}} we
show the reduction of the $B=0$ case to a Newtonian equation discussed above. 
In \textbf{Section \ref{sec:generic}} we
discuss tersely the properties of the generic solution of the systems
described in this paper. 
Finally, in \textbf{Section \ref{sec:conclusions}}, we present some
conclusions and an outlook. The paper is completed by 2 terse {\bf Appendices} 
detailing some aspects of the
reduction of the system (\ref{general}) to its canonical form, as treated in {\bf Section \ref{sec:canon}}.

\section{The problem and its canonical form}
\label{sec:canon}
Clearly, for the purposes we are considering, it makes little difference
whether we consider the system (\ref{general}) for given values of the
parameters $c_{nj}$ or any other system obtained from it by a linear
transformation with time-independent coefficients, such as 
\begin{subequations}
\label{trans}
\begin{eqnarray}
x_{1}\left( t\right) &=&a_{11}y_{1}\left( t\right) +a_{12}y_{2}\left( t\right)
,  \label{transa}
\\
x_{2}\left( t\right) &=&a_{21}y_{1}\left( t\right) +a_{22}y_{2}\left( t\right)
.  \label{transb}
\end{eqnarray}
\end{subequations}
(Note that a rescaling of the time variable by a factor $\lambda $ is
equivalent to a rescaling of $x_{1}$ and $x_{2}$ by the same factor---since
the system (\ref{general}) is invariant under the transformation 
$x_{n}\left( t\right) \Longrightarrow \lambda x_{n}\left( \lambda t\right)$). 
Viewing two systems that can be transformed into one another by such a linear 
transformation as equivalent, we search for a canonical form such that each 
equivalence class contains at most a finite number of
systems in the canonical form. It would, of course, be ideal to have only one
canonical representative in each class, but this does not occur for any of 
the canonical forms we have considered. 
Nevertheless, the identification of a canonical form has two  advantages: first, 
it decreases the number of free parameters, and second it avoids
having to check on a case-by-case basis whether two apparently different 
solvable cases are in fact equivalent in this sense. The canonical 
form given by (\ref{nor}) satisfies the above requirements.
It features only the two  parameters $A$ and $B$. It turns out that every 
generic system (\ref{general}) can be transformed into the form (%
\ref{nor}), but that the corresponding
transformation generically arises in six different ways, so that there exist 
six  linear changes of dependent variables, see (\ref{trans}),
transforming a system from the form (\ref{nor}) into another,
different, such system. Moreover, changing the sign of $x_{1}$ clearly
leaves $A$ invariant but changes the sign of $B$, see (\ref{nor}), so
that these six canonical forms come in pairs featuring equal 
values of $A$ and values of $B$ of opposite sign. The general
formulas connecting the different  canonical forms are given in 
{\bf Appendix \ref{app:b}}.

On the other hand, there do exist  specific systems which cannot be
transformed into the  canonical form (\ref{nor}). These can be
put in the forms 
\begin{subequations}
\label{exc1}
\begin{eqnarray}
\dot{x}_{1}&=&x_{1}x_{2},  \label{exc1a}
\\
\dot{x}_{2}&=&x_{1}^{2}+Bx_{1}x_{2},  \label{exc1b}
\end{eqnarray}
\end{subequations}
or 
\begin{subequations}
\label{exc2}
\begin{eqnarray}
\dot{x}_{1}&=&x_{1}x_{2},  \label{exc2a}
\\
\dot{x}_{2}&=&Ax_{2}^{2}+\sigma x_{1}x_{2},  \label{exc2b}
\end{eqnarray}
\end{subequations}
with $\sigma=0$ or $1$. Note that systems (\ref{exc1}, \ref{exc2}) 
only depend on {\em one\/} parameter, so they are in fact more special 
than the systems of type (\ref{nor}). Finally the form 
\begin{subequations}
\label{exc3}
\begin{eqnarray}
\dot{x}_{1}&=&x_{1}x_{2},  \label{exc3a}
\\
\dot{x}_{2}&=&Bx_{1}x_{2},  \label{exc3b}
\end{eqnarray}
\end{subequations}
while being just the special case $A=0$ of (\ref{nor}), deserves to be
singled out because it corresponds to the only model in the Garnier subclass 
\cite{garnier1,garnier2} which features a free parameter (see below). 
The solution is elementary and given in \cite{kamke};
all the other cases in the Garnier subclass \cite%
{garnier1,garnier2} correspond to fixed values of both $A$ and $B$.

There also exist four uncoupled exceptional forms---which need not be
considered hereafter, since their solution is quite trivial---reading 
\begin{subequations}
\label{uncoupled}
\begin{eqnarray}
\dot{x}_{1}&=&\sigma _{1}x_{1}^{2},  \label{uncoupleda}
\\
\dot{x}_{2}&=&\sigma _{2}x_{2}^{2},  \label{uncoupledb}
\end{eqnarray}
\end{subequations}
where $\sigma _{1,2}$ are either 0 or 1.

In order not to interrupt the flow of the following presentation the proofs
of these results are all relegated to {\bf Appendix \ref{app:a}} (to be read after
the main body of the paper).

The cases having a singlevalued solution described by Garnier 
\cite{garnier1, garnier2} are, on the one hand, among the {\em exceptional\/}
systems described above, 
the system (\ref{exc1}) for $B=0$, called $s_{II}$ by Garnier in \cite{garnier2},
and the system (\ref{exc2}) for $A=1$ and $\sigma=0$, denoted by Garnier as $S_I^2$.  
Four additional non-exceptional cases are listed in Table \ref{tab:1}. 
The first has a solution given in terms of powers and exponential functions, 
whereas the last three have
a solution given in terms of  elliptic functions. In other terms, all these five cases are
either linearisable or reducible to an elliptic ODE. Let us re-emphasize that
in all these cases except the one corresponding to (\ref{exc3}), 
called $s_{IV}$ by Garnier in \cite{garnier2}, these models do not feature
any free parameters.  It should, however, be emphasized that not all systems
of type (\ref{general}) having the property of being singlevalued for all values of
$t$ can be reduced by {\em linear transformations} alone to the normal 
forms listed below. The reduction as performed by Garnier occasionally 
requires nonlinear transformations such as birational transformations.

\begin{table}[tbp]
\begin{center}
\begin{tabular}{ | c | c | c | c |}
\hline
$A$ & $B$  & Garnier list & solution type \\
\hline
$0$ & arbitrary & $s_{IV}$ & linearisable\\ \hline
$-1/2$ & $0$ & $s_V$ & elliptic \\ \hline
$-1$ & $0$ & $s_{VI}$ & elliptic \\
$-1/3$ & $\pm \mathbf{i}\sqrt{2}/3$ & & \\ \hline
$-1/2$ & $\pm\mathbf{i}$ & $s_{VII}$ & elliptic \\
$-1/5$ & $\pm \mathbf{i}/5$ & & \\
$-1$ & $\pm \mathbf{i}$ & & \\ \hline
\end{tabular}
\end{center}
\caption{
Values of $A$ and $B$ for which the system (\ref{nor}) has a singlevalued solution. 
The third column states the name in the article by Garnier \cite{garnier2} and the fourth 
gives the nature of the general solution. Note that the system $s_{IV}$ simply corresponds to 
the system (\ref{exc3}) for $B$ arbitrary. The various systems put under one single heading 
correspond to different canonical forms arising from each 
other through linear transformations.  Note that these are the forms to which 
the canonical forms stated by Garnier can be reduced via linear transformations. 
There exist further canonical systems with the property of singlevaluedness, but
these can be reduced to the standard forms above using appropriate nonlinear transformations.
}
\label{tab:1}
\end{table}

To summarize: if we perform appropriate linear transformations, see (\ref%
{trans}), we reduce any system of the form (\ref{general}) featuring only 
 singlevalued solutions to one of the four specific cases
listed in Table \ref{tab:1}, or else to a model of the type (\ref{exc1}) 
with $B=0$.  As an example of the usefulness of this
reduction procedure, we may point out that, in \cite{farrin1, farrin2, farrin3}, 10 
algebraically solvable cases of (\ref{general}) were identified (as subcases of more general
solvable systems). Following the approach of {\bf Appendix \ref{app:a}}, all of these were found to reduce to 
one of Cases $s_{IV}$ and $s_{VI}$ shown in Table \ref{tab:1}.

In the following, we limit ourselves to considering systems of the canonical
form (\ref{nor}), and we identify a \emph{countably infinite\/} set of values
of $A$ and $B$ such that these systems are algebraically solvable.

\section{A general approach}
\label{sec:solution}
\subsection{Statement of the results}
\label{subsec:res}
We give below {\em sufficient\/} conditions on the values of $A$ and $B$
for the system (\ref{nor}) to be algebraically solvable, and we
exhibit the corresponding solutions of their initial value problems:

\textbf{Case 3.1}: 
\begin{subequations}
\label{res1}
\begin{eqnarray}
A&=&\frac{n+q-1}{n+q},  \label{res1A}
\\
B&=&\pm \frac{n-q}{n+q}\sqrt{\frac{n+q-1}{nq}}
,  \label{res1B}
\end{eqnarray}
\end{subequations}
where $n$ is an arbitrary strictly positive integer, $n>0$, and $q$ is an arbitrary
nonvanishing complex rational number.

\textbf{Case 3.2}: 

\begin{subequations}
\label{res2}
\begin{eqnarray}
A&=&\frac{n+1}{n},  \label{res2A}
\\
B&=&\pm\frac{2q}{n}\sqrt{\frac{n+1}{q^2-n^2}},  \label{res2B}
\end{eqnarray}
\end{subequations}
where $n$ is again an 
arbitrary strictly positive integer, $n>0$, and $q$ a noninteger
complex rational number.

\textbf{Case 3.3}: 

\begin{subequations}
\label{res3}
\begin{eqnarray}
A&=&\frac{n+1}{n},  \label{res3A}
\\
B&=&\pm\frac{2m}{n}\sqrt{\frac{n+1}{m^{2}-n^2}},  \label{res3B}
\end{eqnarray}
\end{subequations}
where $n$ is again an arbitrary strictly positive integer, $n>0$, and $m$ is
an arbitrary integer of parity different from
that of $n$.

\textbf{Case 3.4}: this is the special case with 

\begin{equation}
A=-2,\qquad B=0,  \label{res4}
\end{equation}%
or equivalently $A=-1/5$ and $B=\pm \mathbf{i}3\sqrt{6}/10$. 
Its solution is given in terms of the
square root of a Weierstrass elliptic function; this case is
discussed separately in \textbf{Section \ref{sec:newton}}.

Note that, additionally to \textbf{Cases 3.1-3}, all cases arising from them
by the transformations described in \textbf{Appendix \ref{app:b}} are also 
solvable, but as the corresponding expressions of the parameters $A$
and $B$ become somewhat cumbersome, we do not report them explicitly.

\subsection{Proofs of the results}
\label{subsec:proofs}

In the following, we discuss the details of the treatment of the system 
(\ref{nor}) by quadratures, and we identify a set of cases in which the
corresponding solution can be obtained by algebraic operations,
implying that the analytic continuations of the general
solution $x_{1}\left( t\right) $ and $x_{2}\left( t\right) $ of the
dynamical system (\ref{nor}) has only a finite number of singularities of
the type of algebraic branch points, or equivalently
that---considered as functions of complex $t$---they live on
Riemann surfaces featuring a finite number of sheets.

Define 
\begin{equation}
u\left( t\right) =x_{1}\left( t\right) /x_{2}\left( t\right) ~.  \label{u}
\end{equation}%
From (\ref{nor}) one finds 
\begin{subequations}
\begin{eqnarray}
\dot{u}&=&-x_{2}u(Au^{2}+Bu+A-1)\nonumber\\
&=&-Ax_{2}u(u-u_{+})(u-u_{-}),  \label{equ}
\end{eqnarray}%
where 
\begin{equation}
u_{\pm }=\frac{-B\pm \sqrt{4A-4A^{2}+B^{2}}}{2A}.  \label{u+-}
\end{equation}%
We now define $R(u)$ (consistently with (\ref{u})): 
\end{subequations}
\begin{equation}
x_{2}=R(u),~~~x_{1}=uR(u).  \label{defR}
\end{equation}%
Substituting (\ref{defR}) into (\ref{normb}) and using (\ref{equ}) yields 
\begin{eqnarray}
\dot{x}_{2} &=&R^{\prime }(u)\dot{u}  \notag \\
&=&-AR^{\prime }(u)R(u)u(u-u_{+})(u-u_{-})  \notag \\
&=&x_{2}^{2}\left[ A(1+u^{2})+Bu\right]  \notag \\
&=&R(u)^{2}\left[ A(1+u^{2})+Bu\right]
\end{eqnarray}%
(note that here and hereafter we use the standard mathematical notation
according to which a prime appended to a function denotes a 
differentiation with respect to the  argument of that function).
From the second and fourth lines of these formulas we clearly get%
\begin{equation}
\frac{d}{du}\left[ \ln R(u)\right] =-\frac{1+u^{2}+Bu/A}{u(u-u_{+})(u-u_{-})}%
.  \label{eqR}
\end{equation}%
We now define $\nu _{0}$ and $\nu _{\pm }$ through the relation 
\begin{subequations}
\begin{eqnarray}
-\frac{1+u^{2}+Bu/A}{u(u-u_{+})(u-u_{-})}&=&\frac{\nu _{0}}{u}-\frac{\nu _{+}}{%
u-u_{+}}-\frac{\nu _{-}}{u-u_{-}},\label{defnu}\\
\nu _{0}&=&-1+\nu _{+}+\nu _{-}.  \label{nurel}
\end{eqnarray}
\end{subequations}
Here (\ref{nurel}) follows from (\ref{defnu}) by 
matching the asymptotic behavior as $u\rightarrow \infty $ of the
right-hand side and the left-hand side of (\ref{defnu}).

We thus obtain for $R(u)$, see (\ref{eqR}) and (\ref{defnu}),
\begin{equation}
R(u)=Cu^{\nu _{+}+\nu _{-}-1}(u-u_{+})^{-\nu _{+}}(u-u_{-})^{-\nu _{-}}.
\label{solR}
\end{equation}%
Here $C$ is an integration constant depending on the initial condition,
which enforces the relation 
\begin{subequations}
\begin{equation}
x_{2}(0)=R\left[ u(0)\right]
\end{equation}%
or equivalently 
\begin{eqnarray}
x_{2}(0)^{2}&=&Cx_{1}(0)^{\nu _{+}+\nu _{-}-1}\left[ x_{1}(0)-u_{+}x_{2}(0)%
\right] ^{-\nu _{+}}\times
\nonumber\\
&&\qquad \left[ x_{1}(0)-u_{-}x_{2}(0)\right] ^{-\nu _{-}}.
\label{defC}
\end{eqnarray}
\end{subequations}
The time-dependence of $u$ is obtained by inserting the \emph{ansatz\/} (\ref%
{defR}) into (\ref{equ}): 
\begin{eqnarray}
\dot{u}&=&-AR(u)u(u-u_{+})(u-u_{-})\nonumber\\
&=&-ACu^{\nu _{+}+\nu _{-}}(u-u_{+})^{-\nu_{+}+1}(u-u_{-})^{-\nu _{-}+1},  \label{utime}
\end{eqnarray}%
as stated in \cite{garnier2}.
This integrates to 
\begin{subequations}
\label{solquad}
\begin{eqnarray}
ACt &=&-\int_{u(0)}^{u\left( t\right) }du\,u^{-\nu _{+}-\nu_-}(u-u_{+})^{\nu _{+}-1}_{-}
\times\nonumber\\
&&\qquad(u-u_{-})^{\nu _{-}-1}.
\label{solquada}
\end{eqnarray}
 As stated in the Introduction, we aim to look for values of $A$ and $B$ such that
the connection between $u$ and $t$ given by (\ref{solquada}) is algebraic, that is, that the right hand-side
of (\ref{ACt}) be a rational function of $u^{1/m}$ for some $m\in\mathbb{N}$. 

Further evaluating the above integral we obtain:
\begin{widetext}
\begin{eqnarray}
ACt&=&\int_{u(0)^{-1}}^{u\left( t\right) ^{-1}}dw\,(1-u_{+}w)^{\nu
_{+}-1}(1-u_{-}w)^{\nu _{-}-1} 
\label{solquadb}
\\
&=&\left( 1-\frac{u_{+}}{u_{-}}\right) ^{\nu _{+}}\left( 1-\frac{u_{-}}{u_{+}%
}\right) ^{\nu _{-}}\left[ B_{\rho \left( t\right) }(\nu _{+},\nu
_{-})-B_{\rho (0)}(\nu _{+},\nu _{-})\right] , 
\label{solquadc}
\\
\rho \left( t\right)&=&\frac{u_{-}\left[ u_{+}-u\left( t\right) \right] }{%
u\left( t\right) ~\left( u_{+}-u_{-}\right) }.
\label{solquadd}
\end{eqnarray}
 \label{ACt}
\end{widetext}
\end{subequations}
Here $B_{\rho }(p,q)$ is the  incomplete beta function (see for
instance page 87 of \cite{erd}): 
\begin{eqnarray}
B_{\rho }(\nu _{+},\nu _{-})&=&\int_{0}^{\rho }dt\,t^{\nu _{+}-1}(1-t)^{\nu
_{-}-1}\nonumber\\
&=&\frac{\rho ^{\nu _{+}}}{\nu _{+}}F(\nu _{+},1-\nu _{-};\nu
_{+}+1;\rho )  \label{defIncompleteBeta}
\end{eqnarray}%
where $F(a,b;c;x)$ is the hypergeometric function \cite{erd}. It follows,
from the fact that $F(a,b;c;x)$ is a  polynomial in $x$ whenever
either $-a$ or $-b$ is a non-negative integer (see page 57 of \cite%
{erd}), that the right-hand side of (\ref{solquadb}) is an algebraic expression\/ in $u$ (see
(\ref{solquadd})) whenever $\nu _{-}$ is a strictly
positive integer and $\nu _{+}$ is a (possibly complex) rational number.

Alternative cases in which the right-hand side of (\ref{ACt}) is also an 
algebraic expression in $u$  (not necessarily a polynomial)
can be identified by taking advantage of
the following  three  identities (see eqs. (4), (21), and (22) in Section 2.8 of 
\cite{erd}), 
\begin{widetext}
\begin{subequations}
\label{FHyper}
\begin{eqnarray}
F(a,b;b;z)&=&(1-z)^{-a},  \label{FHypera}
\\
(a)_{m}z^{a-1}F(a+m,b;c;z)&=&\frac{d^{m}}{dz^{m}}\left[ z^{a+m-1}F(a,b;c;z)%
\right] ,  \label{FHyperb}
\\
(c-n)_{n}z^{c-1-n}F(a,b;c-n;z)&=&\frac{d^{n}}{dz^{n}}\left[ z^{c-1}F(a,b;c;z)%
\right] .  \label{FHyperc}
\end{eqnarray}
\end{subequations}
\end{widetext}
It is indeed easily seen that they imply that $F(a+m,b;b-n;z)$ is  
algebraic in $z$ provided $a$ is a   (possibly complex) rational number and both 
$m$ and $n$ are arbitrary nonnegative integers. Note that, in accordance 
to our definition of {\em algebraically solvable}, we do not view irrational powers as algebraic.
Thus---see the
right-hand side of (\ref{defIncompleteBeta})---now we look for those cases
in which 
\begin{subequations}
\label{v+-}
\begin{eqnarray}
\nu _{+}&=&a+m,  \label{v+-a}
\\
1-\nu _{-}&=&b,  \label{v+-b}
\\
\nu _{+}+1&=&b-n,  \label{v+-c}
\end{eqnarray}
\end{subequations}
with $m$ and $n$ arbitrary nonnegative integers. These relations imply
that $\nu _{+}+\nu _{-}$ is a {\em strictly\/}
negative integer or zero, indeed, as seen by subtracting eq. (\ref{v+-b}) from eq. (%
\ref{v+-c}),
\begin{equation}
\nu _{+}+\nu _{-}=-n.  \label{nu+plusnu-}
\end{equation}%
However, see (\ref{eqABnua}) below, the case in which $\nu_++\nu_-=0$
does not arise from any finite value of $A$, and we thus discard it.
Moreover these relations require that $\nu _{+}$ is neither a negative
integer nor zero, since in that case the corresponding
integral in (\ref{solquadb}) has a singularity
of logarithmic type.
Additionally, both $\nu _{+}$ and 
$\nu _{-}$ must be {\em rational}, for which it is enough to assume that,
say, $\nu _{+}$ be rational, see (\ref{nu+plusnu-}).

Let us now express $A$ and $B$ in terms of $\nu _{+}$ and $\nu _{-}$.
Clearly (see (\ref{defnu}) and (\ref{u+-})) 
\begin{subequations}
\begin{eqnarray}
\nu _{+}&=&\frac{1}{Au_{+}(u_{+}-u_{-})},
\\
\nu _{-}&=&-\frac{1}{Au_{-}(u_{+}-u_{-})}.
\end{eqnarray}
\end{subequations}
From this follows 
\begin{subequations}
\label{eqABnu}
\begin{eqnarray}
\nu _{+}+\nu _{-}&=&\frac{1}{1-A},
\\
\nu _{+}\nu _{-}&=&\frac{A}{(1-A)(4A-4A^{2}+B^{2})},\\
A&=&\frac{\nu _{+}+\nu _{-}-1}{\nu _{+}+\nu _{-}},  \label{eqABnua}
\\
B^2&=&\left(
\frac{\nu_+-\nu_-}{\nu_++\nu_-}
\right)^2\frac{\nu_++\nu_--1}{\nu_+\nu_-}
.
\label{eqABnub}
\end{eqnarray}
\end{subequations}
Summarizing, we have at least two distinct cases in which the expression for $t$ (see
(\ref{ACt})) is algebraic in $u$: 
\begin{enumerate}
\item
When one of the two $\nu $'s
is a strictly positive integer and the other is a non-zero
(possibly complex) rational number. This translates into the case
described in (\ref{res1}), if we set $\nu_+=n$
and $\nu_-=q$. 

\item When $\nu _{+}+\nu _{-}$ is a strictly
negative integer, neither $\nu _{+}$ nor $\nu _{-}$ is a negative
integer or zero, and both $\nu _{+}$ and $\nu _{-}$ are non-zero
(possibly complex) rational numbers. This translates into the two possible
cases (\ref{res3}) and  (\ref{res2}) described in {\bf
Subsection \ref{subsec:res}}  
above, depending on whether 
$\nu _{+}-\nu _{-}$ is or is not an integer. In the latter case, we set
$\nu_++\nu_-=-n$ and $\nu_+-\nu_-=q$, and the conditions on $\nu_\pm$
not being a negative integer or zero are automatically fulfilled. If $\nu_+-\nu_-$
is an integer,  $\nu_+-\nu_-=m$,
we note that $m$ and $n$ must have opposite parity for the conditions
on $\nu_\pm$ to be fulfilled, leading to the case shown in (\ref{res3}).  
\end{enumerate}

Let us give  three  examples, chosen to be typical of the three \textbf{Cases 3.1-3},
where the first two belong to {\bf Case 3.1}, the second to {\bf Case 3.2}, and the third
to {\bf Case 3.3}.
In general the solutions are given in terms of two integration constants, $C$ and $t_0$, 
which are determined by the initial conditions: $C$ is given always by (\ref{defC}) whereas 
$t_0$ is determined by substituting $t$ by $0$ and $u$ by $u(0)$ in the relationship 
connecting $u$ and $t$. In general, it is given by:
\begin{equation}
t_0=-\frac{1}{AC}\left( 1-\frac{u_{+}}{u_{-}}\right) ^{\nu _{+}}\left( 1-\frac{u_{-}}{u_{+}%
}\right) ^{\nu _{-}}B_{\rho (0)}(\nu _{+},\nu _{-}) , 
\label{eqt0}
\end{equation}
where $\rho(0)$ is the expression for $\rho$ defined in (\ref{solquadd}), where $u$ is replaced by $u(0)$.

 1) $\nu_-=1$ and $\nu_+$ an arbitrary complex rational number. 
From (\ref{eqABnu})
we find the relationship $B=2A-1$ and $\nu_+=A/(1-A)$. 
From (\ref{ACt}) and (\ref{solR}) we obtain the following expression for
the time dependence of $u$ and the function $R(u)$: 
\begin{subequations}
\label{nup=1}
\begin{eqnarray}
AC(t-t_{0}) &=&
-
\left(
1+\frac{A-1}{Au}
\right)^{A/(1-A)}
\label{nup=1a} \\
R(u) &=&C\frac1{1+u}\left(
1+\frac{A-1}{Au}
\right)^{-A/(1-A)}
\label{nup=1b}
\end{eqnarray}%
\end{subequations}
These define an algebraically solvable system whenever $A$ is a complex rational number.

Somewhat atypically, these relations can be solved  for all $A$, to yield explicit expressions
for $u$ as a function of $t$, and hence for $x_1$ and $x_2$ through (\ref{defR}). Thus:
\begin{subequations}
\begin{eqnarray}
u&=&\frac{A-1}{A}\left\{
[AC(t_0-t)]^{(1-A)/A}-1
\right\}^{-1}\label{solnu=1a}\\
x_1(t)&=&uR(u)\label{solnu=1b}\\
x_2(t)&=&R(u)\label{solnu=1c}
\end{eqnarray}
\label{solnu=1}
\end{subequations}
The solution can therefore be said to extend naturally to irrational values of $A$, though (\ref{nup=1})
is then, of course, no more an algebraically solvable model. Note that, to obtain the full
solution of the initial value problem, it is enough to substitute (\ref{solnu=1a}) into (\ref{solnu=1b}) 
and (\ref{solnu=1c}) using for $R(u)$ the expression in (\ref{nup=1b}).

Note 
in passing that this result can also be obtained directly from (\ref{nor}): the equality,
\begin{equation}
z=x_1+x_2=(1+u)R(u)=\frac{1}{A(t_0-t)}
\end{equation}
which directly follows from (\ref{nup=1}), also follows from (\ref{nor}) through 
the easily verified relation
\begin{equation}
\dot z=Az^2.
\end{equation}
(\ref{norma}) then yields for $x_1$
\begin{equation}
\dot{x}_1=x_1\left[
\frac{1}{A(t_0-t)}-x_1
\right],
\end{equation}
which is a Bernoulli equation linearized by the transformation \cite{ODE} $r=1/x_1$.

2) $\nu _{+}=(1-A)^{-1}-2$ and $\nu _{-}=2$, for arbitrary rational values of $A$. 
From (\ref{eqABnu}) we find that 
and $B=(4A-3)\sqrt{A/(4A-2)}$. From (\ref{ACt}) and (\ref{solR}) we obtain the following expression for
the time dependence of $u$ and the function $R(u)$:

\begin{widetext}
\begin{subequations}
\begin{eqnarray}
\left[
\frac{A^2u_+^2 A(2 A-1)C}{A-1 }(t-t_0)
\right]^{A-1}
&=&u^A
   (u-u_+)^{-2A+1}
   \left\{
u\left[
(A-1)u_- +Au_+
\right]+(1-2A)
u_- u_+ \right\}^{A-1}
\label{ex2a}\\
R(u)&=& u^{A/(1-A)}(u-u_-)^2
   (u-u_+)^{(2A-1)/(1-A)}
   \label{ex2b}
\end{eqnarray}
\label{ex2}
\end{subequations}
\end{widetext}
and thus finally the full solution is given by (\ref{defR}).
Here note that the solution can, in fact, always be obtained in algebraic
terms whenever $A$ is rational. However, the complexity of
the problem increases as the denominator of $A$ grows, and the solution for
irrational $A$ cannot be obtained in elementary algebraic terms. 

3) $\nu _{+}=-7/5$ and $\nu _{-}=2/5$. This corresponds to $A=2$ and 
$B=9/\sqrt{7}$, see (\ref{eqABnu}). We similarly obtain for the time 
dependence of $u$ and the
function $R(u)$: 
\begin{widetext}
\begin{subequations}
\begin{eqnarray}
2C(t-t_0)&=&\frac{7^{4/5} \left(14 u^2+9
   \sqrt{7}\,u+7\right) \left[
   \left(
   9\sqrt{7}\,u+7^{3/5}
   \right)
  w-7 \sqrt{7}\,
   u-7
   \right]}{2^{2/5}\left(2 u+\sqrt{7}\right)^{3/5}
   \left(7 u+\sqrt{7}\right)^{12/5}
   w}\\
   w&=& \left(\frac{5}{\sqrt{7}\,
   u+1}+2\right)^{2/5}\\
   R(u)&=&C\frac{\left(u+\sqrt{7}/2\right)^{2/5}}{u^2
   \left(u+1/\sqrt7\right)
   ^{7/5}}
\end{eqnarray}%
\end{subequations}
\end{widetext}
and again the full solution is
given by (\ref{defR}).

4) $\nu _{+}=1/2$ and $\nu _{-}=-5/2$. This corresponds to $A=3/2$ and 
$B=3\sqrt{3/5}$, see (\ref{eqABnu}). We similarly obtain for the time 
dependence of $u$ and the
function $R(u)$ 
\begin{widetext}
\begin{subequations}
\begin{eqnarray}
\left[
\frac32C(t-t_0)
\right]^2&=&
-5^3 
\left(3
   u-\sqrt{15}
   \right)\frac{\left(123 \sqrt{15}\,
   u^2-180 u+5 \sqrt{15}\right)^2}{16
   \left(15
   u-\sqrt{15}\right)^5}\\
   R(u)&=&C
   \frac{\sqrt{u-\sqrt{5/3}}}{u
   ^3
   \left(u-1/\sqrt{15}\right
   )^{5/2}}
\end{eqnarray}%
\end{subequations}
\end{widetext}
Here again the full solution is given by (\ref{defR}).

In all cases the full evaluation of the solution is reduced to solving for
the zeros of a polynomial, the coefficients of which depend polynomially on
time. The solution thus defines a Riemann surface when $t$ is taken
as a complex variable. The common feature to all these solutions is, of course, 
that they all define Riemann surfaces with a \emph{finite\/} number of sheets.

\section{A Newtonian approach to the case $B=0$}

\label{sec:newton}

We now treat the specific cases of (\ref{nor}) with $B=0$:
\begin{subequations}
\label{special}
\begin{eqnarray}
\dot{x}_{1}&=&x_{1}x_{2},\label{speciala}
\\
\dot{x}_{2}&=&A(x_{1}^{2}+x_{2}^{2}),  \label{specialb}
\end{eqnarray}
\end{subequations}
since it
is possible to treat them in a different and more convenient way. 
First of all, several formulas of \textbf{Section 3} simplify when 
$B=0$; we show how (\ref{eqABnua}) specializes in this case: one then has 
\begin{subequations}
\begin{eqnarray}
\nu _{+}&=&\nu _{-}=\nu ,
\\
A&=&\frac{2\nu -1}{2\nu },
\\
u_{\pm }&=&\pm \sqrt{\frac{1-A}{A}}=\pm \frac{1}{\sqrt{2\nu -1}}.  \label{nu}
\end{eqnarray}
\end{subequations}
Finally (\ref{ACt}) and (\ref{solR}) simplify to 

\begin{subequations}
\label{solquadB=0}
\begin{eqnarray}
ACt &=&\int_{u(0)^{-1}}^{u^{-1}}dw\,\left( 1+\frac{A-1}{A}w^{2}\right) ^{\nu
-1}  \notag \\
&=&\int_{u(0)^{-1}}^{u^{-1}}dw\,\left( 1-\frac{w^{2}}{2\nu -1}\right) ^{\nu
-1},  \label{solquadB=0a} \\
R(u) &=&Cu^{2\nu -1}\left( u^{2}+\frac{A-1}{A}\right) ^{-\nu }  \notag \\
&=&Cu^{2\nu -1}\left( u^{2}-\frac{1}{2\nu -1}\right) ^{-\nu }.
\label{solquadB=0b}
\end{eqnarray}
\end{subequations}
The full list of values of $A$ and $\nu$ which lead to {\em algebraic\/} solutions of (\ref{special})
due to the above arguments are 
\begin{subequations}
\label{intnu}
\begin{eqnarray}
A&=&n/(n+1),\qquad\nu=n,\qquad n\in\mathbb{N}.
\label{intnuplus}\\
A&=&2n/(2n-1),\quad\nu=-(2n-1)/2,\quad n\in\mathbb{N}.
\label{intnuneg}
\end{eqnarray}
\end{subequations}

\textbf{Remark 4.1}. It is interesting to note that none of the values
listed in Garnier \cite{garnier1,garnier2} correspond to a \textit{strictly
positive integer} or \textit{negative half-integer} value of $\nu $. The
solutions $x_{1}\left( t\right) ,$ $x_{2}\left( t\right) $ listed by Garnier
are either elliptic functions, or
correspond to degenerate systems of types (\ref{exc1}) or (\ref{exc2}). 
None correspond to the cases treated in this paper. $\blacksquare $

The system (\ref{special}) allows for an elementary transformation
to a system having the form of the one-dimensional Newtonian equation
(``acceleration equals force'') with a power-law force. Indeed, if we perform
the substitution 
\begin{equation}
x_{1}=z^{-1/A},  \label{subst}
\end{equation}%
we obtain from (\ref{special})
\begin{equation}
\ddot{z}=-A^{2}z^{k},\qquad k=1-2/A.  \label{newton}
\end{equation}

\textbf{Remark 4.2}. Note that via an obvious constant
rescaling of the dependent variable $z$ or of the independent variable $t$
this ODE can be replaced by the ODE (\ref{SimNew}). $\blacksquare $ 

This second-order ODE can clearly be integrated by quadratures (by
multiplying it by $\dot{z}$ and then integrating it over time). Carrying
this out, one obtains expressions a bit simpler than, but essentially
equivalent to, those obtained in \textbf{Section \ref{sec:solution}}. It follows that, if $%
A $ has the values described in (\ref{intnuplus}) and (\ref{intnuneg}),
solving the Newtonian equation (\ref{newton}) with the corresponding values
of $k$ can be reduced to solving a  polynomial equation. The
corresponding values of $k$ and $\nu$ are 
\begin{subequations}
\label{kkk}
\begin{eqnarray}
k&=&-(2n+1)/(2n-1),\qquad \nu=n,\qquad n\in\mathbb{N}, \label{nuplus}
\\
k&=& -(n-1)/n,\qquad \nu=-(2n-1)/2,\qquad n\in\mathbb{N},
\label{numinus}
\end{eqnarray}
\end{subequations}
see (\ref{intnu}).

On the other hand, it follows from the work of Picard \cite{picard} that $%
z(t)$ is  meromorphic in $t$ only for $k$ a nonnegative
integer less than $4$, $k=0,1,2,3$. $k=1$ does not correspond to any finite
value of $A$, whereas the other  three  values correspond to $A=2$, $-2$, and $%
-1$. We saw above (see Table \ref{tab:1} in \textbf{Section  \ref{sec:canon} }) that $A=-1$ is a 
case in which the solution is 
singlevalued, as discussed by Garnier \cite{garnier1}; while the case $A=2$
corresponding to $\nu =-1/2$ was discussed above, see (\ref{intnuneg}).

But the case $A=-2$ is  {\em new}: specifically it corresponds to $\nu
=1/6 $, which does not fall in the cases discussed in \textbf{Section  \ref{sec:solution}},
nor does it belong to the Garnier list. This case corresponds to $k=2$. The
solution $z\left( t\right) $ of (\ref{newton}) is an elliptic
function, but for $x_{1}(t)$ we have
\begin{subequations}
\label{A=-2}
\begin{eqnarray}
x_{1}(t) &=&z(t)^{1/2},  \label{A=-2a} \\
z(t) &=&-\exp(2\pi \mathbf{i}/3)\left( \frac{3}{2}\right) ^{1/3}\times\nonumber\\
&&\qquad
\wp \left[ \left( -%
\frac{2}{3}\right) ^{1/3}\!\left( t+C_{1}\right) ;0,C_{2}\right] ,
\label{A=-2b}
\end{eqnarray}%
\end{subequations}
where $C_{1}$ and $C_{2}$ are integration constants, and where $\wp
(x;g_{2},g_{3})$ denotes the Weierstrass function \cite{erd} as a function
of its \emph{invariants\/} $g_{2}$ and $g_{3}$. $x_{1}(t)$ is thus 
not singlevalued, but can be obtained from the singlevalued function 
$z\left( t\right) $ by an algebraic operation, namely by taking the 
square root.

\textbf{Remark 4.3}. Note that the case $A=-1$ mentioned two paragraphs
above is similar, but in this case the connection between $z(t)$
and $x_{1}(t)$ introduces no loss of analyticity, so that the
result is in the class of {\em singlevalued\/} solutions, and it indeed appears in
the list of Garnier.

\textbf{Remark 4.4.} The fact that the solutions of the simple Newtonian
equation (\ref{newton}) for all the assignments (\ref{kkk}) of $k$ are 
algebraic functions of $t$ does not seem to have been noticed
earlier; note that this implies that all corresponding, 
appropriately modified, systems are isochronous (as detailed at the
end of {\bf Section  \ref{sec:intro}}, see (\ref{NewtIso})). On the other hand it seems likely that for all
sufficiently large positive integer values of the exponent $k$ in (%
\ref{newton}) the solution of this simple Newtonian equation is 
not algebraic, leading---in the complex---to extremely complicated
behavior (see \cite{GS2007} for a detailed treatment when $k$ is a 
strictly positive even number). $\blacksquare$

\section{Qualitative properties of the generic solution}

\label{sec:generic}

In the following we consider the qualitative properties of
the systems treated in \textbf{Section  \ref{sec:solution}} 
when they start from generic initial
conditions. Such initial conditions are, of course, 
complex. To be specific, we concentrate on the \textbf{Cases 3.1,
3.2} and \textbf{3.3} described in {\bf Subsection 
\ref{subsec:res}}.  We shall, in the
following, always consider the solution's behavior for
\emph{real\/} times. The first finding we report is the
following: for generic initial conditions the
solution remains bounded for all finite
real times. In other words, the solution never
blows up at a finite time. Moreover, this generic
solution remains analytic for all
finite times, i.e. it never hits a singularity. 
This is seen as follows: $u(t)$, or some 
algebraic function of it, is a zero of some 
$t$-dependent polynomial. A first possible way in which a singularity might
arise, is if the coefficients of the $t$-dependent polynomial at some time
take values such that the polynomial has a 
multiple zero. When this happens, the discriminant of the
polynomial must vanish, which entails \emph{two\/} real conditions.
If the time is real, generically the curve in the space of
polynomials will therefore not hit the set of polynomials with
multiple zeros, since this set has real codimension two  in the set of
all polynomials. Another possibility is that at some point $u(t)$
takes a value for which the denominator of $R(u)$ vanishes, thus leading to
the divergence of $x_{2}(t)$. There are only three  such values, however, namely $%
0$ and $u_{\pm }$, see (\ref{solR}). Again, going through one of these 
values corresponds to two real conditions and will therefore generically not happen.

What can we say concerning the behavior of $x_{1}(t)$ and $x_{2}(t)$ for
large times? From (\ref{solquada}) we see that $u$
must approach one of the three values $0$ or $u_{\pm }$ (going to infinity
is not an option for $u$, as the integral in $u$ in (\ref{solquada}) 
converges as $u\rightarrow \infty $, since the integrand goes as $%
u^{-2}$). Under these circumstances, since $u$ is defined 
as the root of a polynomial
equation, as $t$ diverges, it remains close to a fixed value; it will hence eventually---for 
$t$ sufficiently large---not move from one branch to another, and thus tend
monotonically to a given value. For similar reasons, the function $R(u)$, see (\ref{solR})
will tend to a value which is either 0 or infinity, with a power-law that
can be determined in each specific case.

\begin{figure}[tbp]
\begin{center}
\includegraphics[scale=0.75]{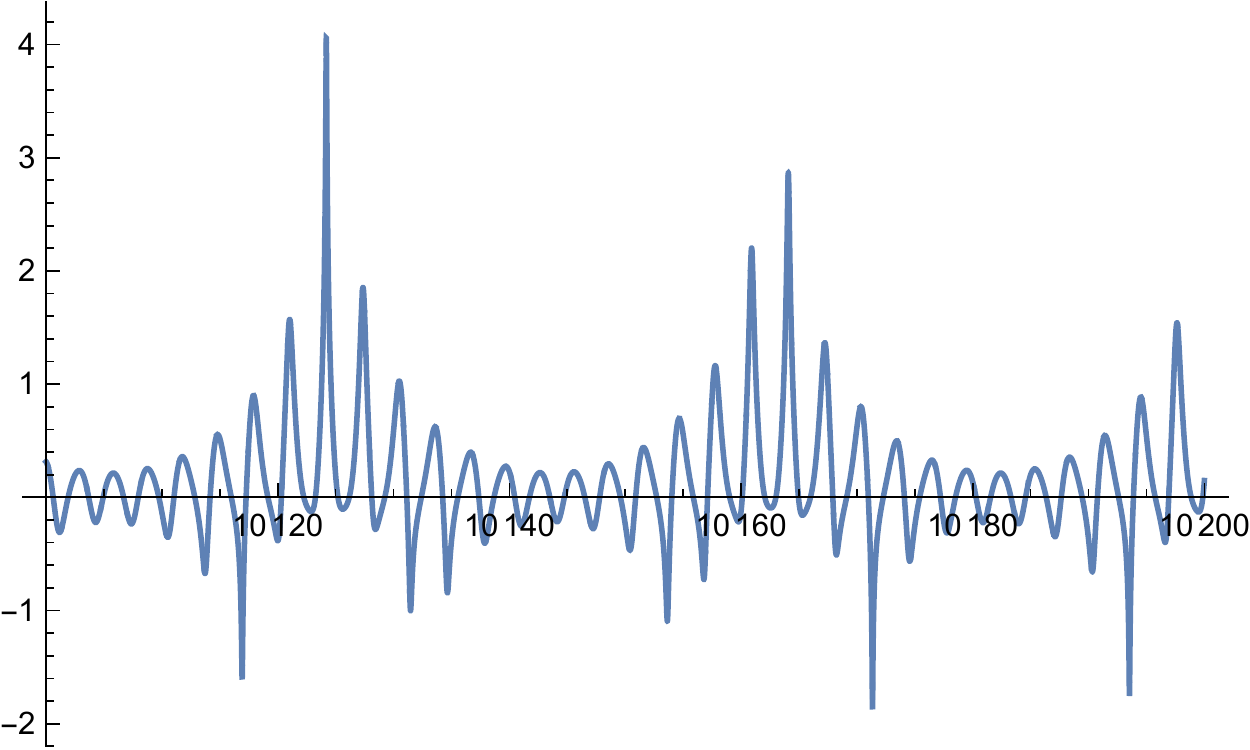}
\end{center}
\caption{ Plot of  $\ln[|x_1(t)|^2]$ for the system (\ref{nor}) with $A=-2$ and $B=0$
for large times $t\approx10000$.
Note how the function does not have a simple behavior, but rather shows sharp peaks.}
\label{fig:13}
\end{figure}

Such smoothness properties are by no means obvious.
For instance they fail in the
case of arbitrary real $A$ and $B$ and real
initial conditions: indeed in this case, it is well known, and also readily verified
from the results shown in {\bf Section  \ref{sec:solution}}, that 
the solutions of (\ref{nor}) can  diverge at 
finite time for an open set of initial
conditions. The result similarly fails in the case $A=-2$ and $B=0$, see (\ref{A=-2b}).
The solutions then do not have the regular behavior at infinity described in the
last paragraph. Indeed, for generic values of $C_{1}$ and $C_{2}$, the poles 
of the elliptic function there given, while they do not typically lie on the $t$ axis,
in general come arbitrarily close to it, since they lie on a lattice.
The function $x_{1}(t)$ thus becomes {\em arbitrarily large infinitely often}, but
irregularly so, as $t\rightarrow \infty $, see Figure \ref{fig:13}. In this
sense the systems whose solutions are described in 
{\bf Section  \ref{sec:solution}}
are therefore remarkably simple in their regularity properties  for finite $t$
as well as in their asymptotic behavior.

\section{Conclusions and outlook}

\label{sec:conclusions}

We have displayed a large variety of special cases of the system 
of ODE's (\ref{nor}) with the property that their solution, 
in the {\em complex\/} $t$-plane, only takes a {\em finite\/} number of different values
according to the path taken, or in other words, that they
define a Riemann surface with only a {\em finite\/} number of sheets. 
These solutions have the additional remarkable
feature of being rather simple: in all cases but one, they can be
calculated from $t$ by purely algebraic operations. They
are therefore significantly simpler than the only other explicit solutions
of (\ref{nor}) previously known, namely those listed by
Garnier, which are elliptic functions. For the one
exceptional case, (\ref{nor}) with $A=-2$ and $B=0$, 
it is sufficient to calculate an elliptic function of $t$
and take its square root. This constitutes in itself a remarkably simple
addition to the list presented by Garnier.

These systems are additional to those already listed by
Garnier, which were obtained by looking for systems whose solution
has no branch points at complex times, hence is a {\em singlevalued\/} function
of the complex variable $t$. Since there were
only five such cases, it is seen that loosening the requirement of being  
singlevalued by allowing the presence of a finite number of 
algebraic branch points significantly increases the number of examples, 
since we have found a doubly infinite set of such models (see 
(\ref{nor}) with (\ref{res1}), (\ref{res2}), (\ref{res3})).
Moreover we have noted that, via the change of dependent and independent
variables (\ref{isotrans}), all these systems yield new autonomous
nonlinearly-coupled dynamical systems featuring the remarkable property to
be isochronous.

As an additional finding, we found that the special case $B=0$ of (\ref{nor}) can be transformed
to the Newtonian equation (\ref{SimpleNewt}), so that our results of Section \ref{sec:solution}
can be extended to the Newtonian equation (\ref{SimNew}) for those values of $k$ stated in 
(\ref{kknn}).  This finding, which implies that the very simple Newtonian equation
(\ref{SimpleNewt}) is algebraically solvable  when the exponent $k$  has one of the infinite series
of values listed in (\ref{kknn}), and likewise that the Newtonian equation (\ref{NewtIso}) 
is isochronous for the same values of $k$, seems to us quite remarkable.

Two interesting open problems are the following: first, we have only given {\em sufficient\/} 
conditions for the algebraic solvability of (\ref{general}). It is clearly of interest to be able to
give a complete list of systems with this property. On the other hand, it would also 
be important to know how our systems behave as far as the existence of invariants is concerned.
Finally, let us point out that Sokolov and Wolf \cite{SW} generalized Garnier's results to the case of 
quadratic systems with non-commuting variables:
it might be an interesting direction of research to search for a similar generalisation of our results.
Another interesting question concerns the possibility of extending this approach to non-homogeneous systems,
in particular such as involve both linear and quadratic terms; for preliminary work in this direction, see
for example \cite{farrin1,farrin2,farrin3}.

\section*{Authors' contributions}

All authors contributed equally to this work.

\section*{Acknowledgments}

One of us (FL) would like to acknowledge financial support by 
UNAM PAPIIT-DGAPA-IN103017 as well as by CONACyT Ciencias B\'asicas 
254515. All authors would like to acknowledge the Centro Internacional de Ciencias
in Cuernavaca, which organized the event {\em Integrable Systems and Beyond\/}
in November--December 2018, where this research project was initiated. 
One of us (FC) would like to thank Paolo Santini for
enlightening suggestions about the identification of the singularities of nonlinear
ODE's.

\section*{AIP Publishing Data Sharing Policy}

Data sharing not applicable---no new data generated.

\appendix

\section{Canonical forms for the system (\protect\ref{general})}

\label{app:a}

We display here the reduction of an arbitrary system of the form (\ref%
{general}) to either the form (\ref{nor}) or one of the exceptional forms
(\ref{exc1}), (\ref{exc2}), and (\ref{exc3}), via a linear transformation of the
two dependent variables, see (\ref{trans}).

From $u=x_{1}/x_{2}$ (see (\ref{u})) we obtain 
\begin{subequations}
\label{uP}
\begin{eqnarray}
\dot{u}&=&-x_{2}P_{3}(u),  \label{uPa}
\\
P_{3}(u)&=&c_{21}u^{3}-(c_{11}-c_{23})u^{2}-(c_{13}-c_{22})u-c_{12}~.
\label{P3u}
\end{eqnarray}
\end{subequations}
We now distinguish  two  main cases with several subcases.

\textbf{Case A1}. Let the polynomial $P_{3}(u)$ have (at least) one simple zero:
the generic case is, of course, that all its three zeros
are simple. In the presence of a double zero, the other
zero is of course simple. In that case we take $\overline{u}$ to be
such a zero. We then define (see (\ref{u})) 
\begin{equation}
y_{1}=x_{1}-\overline{u}x_{2}=\left( u-\bar{u}\right) x_{2},\qquad
y_{2}=x_{2}.  \label{defy1}
\end{equation}%
Because $\overline{u}$ is a zero of $P_{3}(u)$, if $u$ starts out at $%
\overline{u}$, it maintains that value always (see (\ref{uPa})). It follows
that, if $y_{1}$ starts out at zero, it always remains there. The
equation for $y_{1}$ is thus of the form 
\begin{equation}
\dot{y}_{1}=y_{1}(\alpha y_{1}+\beta y_{2}).
\end{equation}%
Since $\overline{u}$ is a simple zero of $P_{3}(u)$, it follows that 
\begin{equation}
\beta \neq 0.
\end{equation}%
We can thus introduce $z_{1}=y_{1}$ and $z_{2}=\alpha y_{1}+\beta y_{2}$ as 
2 new independent variables. One then has 
\begin{equation}
\dot{z}_{1}=z_{1}z_{2},\qquad\dot{z}%
_{2}=A_{1}y_{1}^{2}+A_{2}y_{2}^{2}+By_{1}y_{2}  \label{prenormb}
\end{equation}%
and, by appropriately rescaling $y_{1}$, we obtain (\ref{nor})
if neither $A_{1}$ nor $A_{2}$ vanish (up to the formal exchange of $%
y_{1},y_{2}$ with $x_{1},x_{2}$). Likewise, if either $A_{1}$ or $%
A_{2}$ vanish, we obtain forms (\ref{exc1}) and (\ref{exc2}) respectively; 
and if both $A_{1}$ and $A_{2}$ vanish, we obtain (\ref{exc3}).

\textbf{Case A2.} The polynomial $P_{3}(u)$ is quadratic, i.e. 
$c_{21}=0$, and it features a double zero, implying 
\begin{equation}
(c_{13}-c_{22})^{2}-4c_{23}c_{12}=0.
\end{equation}%
In this case we first try to invert the roles of $x_{1}$ and $x_{2}$. If
this leads to a generic polynomial of third degree, we are led back
to the first case. If not, we have overall
\begin{subequations}
\begin{eqnarray}
c_{21} &=&0, \\
c_{12} &=&0, \\
(c_{13}-c_{22})^{2}-4c_{23}c_{12} &=&0, \\
(c_{23}-c_{11})^{2}-4c_{13}c_{21} &=&0.
\end{eqnarray}
\end{subequations}
But the last  two  equations reduce to 
\begin{subequations}
\begin{eqnarray}
c_{13} &=&c_{22}, \\
c_{23} &=&c_{11},
\end{eqnarray}%
\end{subequations}
which leads to $P_{3}(u)=0$. This system can then be reduced to the
uncoupled forms (\ref{uncoupled}).

\section{Equivalent canonical forms}

\label{app:b}

Let a system in the canonical form (\ref{nor}) have the
parameters $A$ and $B$. Clearly the system with the parameters $A$ and $-B$
is equivalent via a change of sign of $x_{1}$. On the other hand, it can be
seen, using straightforward but tedious calculations, that the following  four 
values are also equivalent:
\begin{subequations}
\label{equiv}
\begin{eqnarray}
A(\sigma _{1},\sigma _{2}) &=&\frac{1}{\Delta }\left[ 4A^{2}-2A-B^{2}-\sigma
_{1}R\right],
\label{equiva} \\
B(\sigma _{1},\sigma _{2}) &=&\frac{\sigma _{2}}{B\Delta }
\bigg\{ 
B^{2}\bigg[%
1-4A(A-1)+B^{2}\bigg]+\nonumber\\
&&\qquad\sigma _{1}\left( 1-4A^{2}+B^{2}\right) R
\bigg\} ,
\label{equivb} 
\\
\Delta &=&2A\left[ (1-2A)^{2}-B^{2}\right],
\label{equivc}
\\
R &=&\sqrt{B^{2}(4A-4A^{2}+B^{2})},
\label{equivd}
\end{eqnarray}%
\end{subequations}
where $\sigma _{1}$ and $\sigma _{2}$ each takes the values $+1$ and $-1$.
These  four  sets of values together with the  two  initial sets of values lead to 
$6$ equivalent canonical forms.

The route to arrive at these
results, see (\ref{equiv}), goes as follows: first the general substitution
\begin{subequations}
\begin{eqnarray}
x_{1} &=&a_{11}y_{1}+a_{12}y_{2}, \\
x_{2} &=&a_{21}y_{1}+a_{22}y_{2}
\end{eqnarray}%
\end{subequations}
is performed in (\ref{nor}) and the resulting equations for the new variables $y_{1}\left(
t\right) $ and $y_{2}\left( t\right) $ are computed. These depend on 
$a_{11} $, $a_{12} $, $a_{21} $ and $a_{22}  $. The conditions stating that these new
equations are again in the canonical form (\ref{nor}) are then determined and solved using
Mathematica and yield
\begin{subequations}
\begin{eqnarray}
a_{11} &=&\sigma _{2}\frac{2A-4A^{2}+B^{2}-\sigma _{1}R}{\Delta }, \\
a_{12} &=&\frac{-B^{2}+(2A-1)\sigma _{1}R}{B\Delta }, \\
a_{21} &=&\sigma _{2}\frac{-B^{2}-(2A-1)\sigma _{1}R}{B\Delta }, \\
a_{22} &=&\frac{\sigma _{1}\sigma _{2}(2A-4A^{2}+B^{2})+\sigma _{1}R}{\Delta 
},
\end{eqnarray}%
\end{subequations}
where again $\sigma _{1}$ and $\sigma _{2}$ each takes the values $+1$ and 
$-1$. Putting these values into the transformed equations yields the result
stated above, see (\ref{equiv}).

\end{document}